\documentclass[aps,pra,preprint,showpacs,letterpaper]{revtex4-1}
\usepackage{CJK}
\usepackage{amsmath}
\usepackage{amsfonts}
\usepackage{bm}
\usepackage{graphicx}
\usepackage{array}
\usepackage{dcolumn}
\usepackage{mathrsfs}
\begin{document}
\title{Investigation of the laser-induced lineshape change in attosecond transient absorption spectra by employing a time-dependent generalized Floquet approach}
\date{\today}
\author{Di Zhao}
\email{d.zhao@mail.xjtu.edu.cn}
\affiliation{School of Physics, Xi'an Jiaotong University, Xi'an, 710049, China}
\affiliation{Shaanxi Province Key Laboratory of Quantum Information and Quantum Optoelectronic Devices}
\author{Chen-Wei Jiang}
\affiliation{School of Physics, Xi'an Jiaotong University, Xi'an, 710049, China}
\affiliation{Shaanxi Province Key Laboratory of Quantum Information and Quantum Optoelectronic Devices}
\author{Ai-Ping Fang}
\affiliation{School of Physics, Xi'an Jiaotong University, Xi'an, 710049, China}
\affiliation{Shaanxi Province Key Laboratory of Quantum Information and Quantum Optoelectronic Devices}
\author{Shao-Yan Gao}
\affiliation{School of Physics, Xi'an Jiaotong University, Xi'an, 710049, China}
\affiliation{Shaanxi Province Key Laboratory of Quantum Information and Quantum Optoelectronic Devices}
\author{Fu-li Li}
\affiliation{School of Physics, Xi'an Jiaotong University, Xi'an, 710049, China}
\affiliation{Shaanxi Province Key Laboratory of Quantum Information and Quantum Optoelectronic Devices}

\begin{abstract}

We introduce a time-dependent generalized Floquet (TDGF) approach to calculate attosecond transient absorption spectra of helium atoms subjected to the combination of an attosecond extreme ultraviolet (XUV) pulse and a delayed few-cycle infrared (IR) laser pulse. This TDGF approach provides a Floquet understanding of the laser-induced change of resonant absorption lineshape. It is analytically demonstrated that, the phase shift of the time-dependent dipole moment that results in the lineshape changes consists of the \emph{adiabatic} laser-induced phase (LIP) due to the IR-induced stark shifts of adiabatic Floquet states and the \emph{non-adiabatic} phase correction due to the non-adiabatic IR-induced coupling between adiabatic Floquet states. Comparisons of the spectral lineshape calculated based on the TDGF approach with the results obtained with the LIP model [S. Chen \emph{et al.}, Phys. Rev. A \textbf{88}, 033409(2013)] and the rotating-wave approximation (RWA) are made in several typical cases. It is suggested in the picture of adiabatic Floquet states that, the LIP model works as long as the generalized adiabatic theorem [A. Dodin \emph{et al.}, Phys. Rev. X Quantum \textbf{2}, 030302(2021)] fulfils, and the RWA works when the higher-order IR-coupling effect in the formation of adiabatic Floquet states is neglectable.
\end{abstract}
\maketitle

\section{Introduction}
In the past decades, the advent of attosecond pulses offers a powerful tool to trace the electronic response in atoms and molecules in atoms \cite{Nat466-739, PRA97-031407, NJP18-013041, PRA87-063413, Sci354-738, PRL109-073601, OL37-2211, PRL106-123601, NP13-472, Sci340-716, Nat516-374}, molecules \cite{PRL121-023203, PRA98-053401, CPL683-408, PRA94-023403, PRA95-043427, PRA91-043408, NPhoton16-196, PRL127-123001, PRA101-023401} and solid targets \cite{Sci346-1348, Sci353-916, Sci357-1134, PRB94-165125, NPhoton16-33, PRL124-207401, PRB104-064103} in the natural timescale of electronic dynamics \cite{RMP81-163, JPB49-062001, NPhoton11-252, NP3-381, ARPC63-447}. An attosecond extreme ultraviolet (XUV) pulse synchronized with an infrared (IR) laser pulse gives rise to a transient absorption measurement, so-called attosecond transient absorption spectroscopy (ATAS), to investigate the IR-laser-driven electronic dynamics at the sub-optical-cycle scale, with the temporal resolution indirectly accessed through the highly controlled time delay between pulses. For atomic systems, typical features of ATA spectra include \cite{JPB49-062003,PRA96-013430}: light-induced states (LISs) and subcycle oscillating fringes when the XUV pulse arrives during the IR pulse, hyperbolic sidebands when the XUV pulse arrives before the IR pulse, and IR-induced change of resonant absorption lineshape for delays in which the XUV pulse precedes or overlaps with the IR pulse.

The Floquet theory has brought a conceptual picture of ATA process \cite{JPB49-062003}. For instance, the LISs in ATA spectra are the Floquet states that are dipole allowed to the ground state, and the subcycle oscillations are the quantum beating between different Floquet states that collapse to the same bare state when the IR pulse turns off \cite{JPB49-062003,PRA92-033408,SR3-1105}. Yet the IR-induced change of resonant absorption lineshape has not been explored in the Floquet formalism. On the other hand, Ott \emph{et al} \cite{Sci340-716} and Chen \emph{et al} \cite{PRA88-033409} propose a laser-induced phase (LIP) model under the lowest-order assumption, in which the laser-induced modification of the population of the XUV-excited bright state is neglected and only a laser-induced modification of the bright state phase is included. The phase shift of the bright state gives rise to a phase shift of the time-dependent dipole moment relative to the XUV-only case, resulting in the laser-induced lineshape changes in ATA spectra. Several potential applications, such as the reconstruction of laser-driven electronic dynamics \cite{Nat516-374,PRL121-173005} and the generation of XUV frequency combs \cite{NJP16-093005}, have been proposed based on the LIP model. However in the frame of the LIP model, neither the second order perturbation theory (SOPT) nor the rotating wave approximation (RWA) is a universal approach to accurately reproduce the IR-induced change of resonant absorption lineshape \cite{JPB49-062003}.

In this paper, we introduce a time-dependent generalized Floquet (TDGF) approach to solve the time-dependent Schr\"{o}dinger equation (TDSE) of XUV-excited helium atom driven by a delayed few-cycle IR pulse, to investigate the IR-induced lineshape change in ATA spectra in the picture of adiabatic Floquet states. By employing the generalized adiabatic theorem \cite{PRXQ2-030302}, it is demonstrated that the lowest-order assumption in the LIP model is equivalent to the generalized adiabatic approximation (GAA) and the LIP is the difference of dynamical phase shifts of adiabatic Floquet states. The total IR-coupling-induced phase shift imported on the time-dependent dipole moment consists of two parts, the \emph{adiabatic} LIP due to the IR-induced stark shifts of adiabatic Floquet states and the \emph{non-adiabatic} phase correction due to the IR-induced non-adiabatic coupling between adiabatic Floquet states. We present comparison of resonant absorption lineshape calculated based on the TDGF approach with those based on GAA and RWA in a He-like few-level model. It is suggested that, the failure of LIP in the case of weak driving and small detuning is due to the non-ignorable IR-induced-coupling between adiabatic Floquet states, i.e. the breakdown of generalized adiabatic theorem, while the failure of RWA in the case of strong driving and large detuning is due to the neglection of higher-order coupling in the formation of adiabatic Floquet states. Our numerical calculation also shows that, the TDGF calculation of resonant absorption lineshape in a He-like few-level model agrees qualitatively with the \emph{ab initio} calculation of single-active-electron (SAE) helium atom.

The paper is organized as follow. First, we briefly describe the single-atom response function for the calculation of ATA spectra in Sec.\ref{sec_RF}, introduce the TDGF approach in Sec.\ref{sec_TDGF}, and derive the LIP by employing the generalized adiabatic approximation in the frame of adiabatic Floquet states in Sec.\ref{sec_LIP}. Then Sec.\ref{sec_dis} presents the investigation of IR-induced lineshape change via the comparison of calculated ATA spectra by employing different approaches, for the cases with weak driving and moderate detuning in Sec.\ref{subsec_wm}, strong driving and large detuning in Sec.\ref{subsec_sl} and weak driving and small detuning in Sec.\ref{subsec_ws}. Finally, in Sec.\ref{subsec_ab}, we examine the efficiency of the TDGF approach via the comparison with the \emph{ab initio} calculation of SAE Helium atom. Summaries and concluding remanks are given in Sec.\ref{sec_con}. Atomic units $(\hbar=e=m_{e}=a_{0}=1)$ are used throughout, unless indicated otherwise.

\section{Theory and methods}
\label{sec_theo_meth}
\subsection{The single-atom response function and TDSE calculation}
\label{sec_RF}
We consider a He-like three-level system driven by the combination of an attosecond XUV pulse and a few-cycle IR pulse. The $1s^2$, $1s2p$ and $1s2s$ states of Helium atom are included and labelled as $|\alpha\rangle$, $|\beta\rangle$ and $|\gamma\rangle$, respectively. The time-dependent wavefunction can be expanded as
\begin{equation}
|\Psi(t)\rangle=C_{\alpha}(t)e^{-i\mathcal{E}^{0}_{\alpha}t}|\alpha\rangle+C_{\beta}(t)e^{-i\mathcal{E}^{0}_{\beta}t}|\beta\rangle+C_{\gamma}(t)e^{-i\mathcal{E}^{0}_{\gamma}t}|\gamma\rangle,
\label{eq:psi_3lvl}
\end{equation}
in which $\mathcal{E}^{0}_{\alpha}$, $\mathcal{E}^{0}_{\beta}$ and $\mathcal{E}^{0}_{\gamma}$ are the eigenenergies of field-free states, and $C_{\alpha}$, $C_{\beta}$ and $C_{\gamma}$ are the probability amplitudes of corresponding states. The time-dependent Hamlitonian is written as
\begin{equation}
\begin{split}
\hat{H}(t)=&\hat{H}_{0}+\hat{V}(t)\\
=&|\alpha\rangle\mathcal{E}^{0}_{\alpha}\langle\alpha|+|\beta\rangle\mathcal{E}^{0}_{\beta}\langle\beta|+|\gamma\rangle\mathcal{E}^{0}_{\gamma}\langle\gamma|\\
&-E^{0}_{X}(t)\cos(\omega_{X} t)(\mu^{0}_{\alpha\beta}|\alpha\rangle\langle\beta|+\mu^{0}_{\beta\alpha}|\beta\rangle\langle\alpha|)\\
&-E^{0}_{I}(t-\tau)\cos (\omega_{I} t-\omega_{I}\tau)(\mu^{0}_{\beta\gamma}|\beta\rangle\langle\gamma|+\mu^{0}_{\gamma\beta}|\gamma\rangle\langle\beta|),
\end{split}
\label{eq:H_3lvl}
\end{equation}
in which $\mu^{0}_{\alpha\beta}=\mu^{0\ast}_{\beta\alpha}=e\langle\alpha|z|\beta\rangle$ and $\mu^{0}_{\beta\gamma}=\mu^{0\ast}_{\gamma\beta}=e\langle\beta|z|\gamma\rangle$ are the transition matrix elements, $\omega_{I(X)}$ and $E^{0}_{I(X)}(t)$ are the carrier frequency and pulse-envelope function of the IR (XUV) pulse, respectively, and the two pulses with the relative time delay $\tau$ are both polarized along the $z$ axis. The XUV pulse couples the $|\alpha\rangle$ and $|\beta\rangle$ states, and the IR pulse only couples the $|\beta\rangle$ and $|\gamma\rangle$ states. The time-dependent dipole moment $d(t)$ is determined by
\begin{equation}
d(t)=C^{\ast}_{\alpha}(t)C_{\beta}(t)\mu^{0}_{\alpha\beta} e^{-i(\mathcal{E}^{0}_{\beta}-\mathcal{E}^{0}_{\alpha})t}+c.c.,
\label{eq:td_dipole_3lvl}
\end{equation}
where we only measure frequencies relative to the $|\alpha\rangle$ state. The coefficients $C_{\alpha}(t)$ and $C_{\beta}(t)$ can be obtained by numerically solving the TDSE with the Hamiltonian Eq.(\ref{eq:H_3lvl}) by employing the fourth-order Runge-Kutta(RK) method.

The dipole moment $d(t)$ that initiated by the interaction with the XUV pulse freely oscillates after the radiation fields have passed, and it would decay due to collisional broadening or spontaneous decay. In order to include these experimental effects in our calculation, the dipole moment $d(t)$ is multiplied by a smoothly decaying function before the Fourier transformation:
\begin{equation}
W(t)=\left\{
{\begin{array}{ll}
e^{-\frac{\Gamma}{2} (t)}&, \quad t>0\\
1&, \quad t\leq0,\\
\end{array}
}\right.
\label{eq:winfunc}
\end{equation}
where $\Gamma$ stands for the phenomenological decay rate. Then we have the Fourier transform of the time-dependent dipole moment $d(t)$,
\begin{equation}
\label{eq:dip_total_omega}
\tilde{d}(\omega)=\mu^{0}_{\alpha\beta} \int_{\tau}^{\infty}C^{\ast}_{\alpha}(t)C_{\beta}(t)e^{i[\omega-(\mathcal{E}^{0}_{\beta}-\mathcal{E}^{0}_{\alpha})]t} W(t)dt,
\end{equation}
and the absorption spectra can be expressed as the response function defined by \cite{PRA11Gaarde}
\begin{equation}
S(\omega)=2\text{Im}[\tilde{d}(\omega)\tilde{E}^{\ast}(\omega)],
\label{eq:ResFun}
\end{equation}
where $\tilde{E}(\omega)$ is the Fourier transform of the total laser field $E(t)$. Note that, although the complete description of transient absorption spectra must account for the reshaping of an XUV pulse during the propagation, the single-atom response yields similar results as the full calculation for the helium atom \cite{PRA11Gaarde,PRA12Chen}.

\subsection{The time-dependent generalized Floquet approach}
 \label{sec_TDGF}
Since the preceding attosecond XUV pulse can be approximated by a $\delta$-function due to its short duration, the three-level system can be further simplified into a primitive two-level system involving only the $|\beta\rangle$ and $|\gamma\rangle$ states and the IR pulse arriving after the XUV pulse. Here we introduce a time-dependent generalized Floquet (TDGF) approach for the evolution of the laser-dressed two-level system, the time-dependent Hamiltonian of which is given by
\begin{equation}
\begin{split}
\hat{H}_{I}(t;E_{t})=&\hat{H}^{\prime}_{0}+\hat{V}_{I}(t)\\
=&|\beta\rangle\mathcal{E}^{0}_{\beta}\langle\beta|+|\gamma\rangle\mathcal{E}^{0}_{\gamma}\langle\gamma|\\
&-\frac{1}{2}E_{t} (e^{i\omega_{I}(t-\tau)}+e^{-i\omega_{I}(t-\tau)})(\mu^{0}_{\beta\gamma}|\beta\rangle\langle\gamma|+\mu^{0}_{\gamma\beta}|\gamma\rangle\langle\beta|)\\
\end{split}
\label{eq:HI}
\end{equation}
with $E_{t}=E^{0}_{I}(t)$ indicating the time-dependent amplitude of the IR pulse. For each value of $E_{t}$, the Hamiltonian $\hat{H}_{I}(t;E_{t})$ is periodic and the instantaneous Floquet quasienergy states can be defined as
\begin{equation}
|\psi_{\lambda}(t;E_{t})\rangle=e^{-i\theta_{\lambda}(t;E_{t})} |\lambda(t;E_{t})\rangle,
\label{eq:TD-Flo-state}
\end{equation}
with the dynamical phase $\theta_{\lambda}(t;E_{t})\equiv \int_{0}^{t}dt^{\prime} \mathcal{E}_{\lambda}(t^{\prime};E_{t^{\prime}})$, in which the instantaneous Floquet quasienergy $\mathcal{E}_{\lambda}(t;E_{t})$ and corresponding quasienergy state $|\lambda(t;E_{t})\rangle$ satisfy the eigenvalue equation
\begin{equation}
\hat{H}_{F}(t;E_{t})|\lambda(t;E_{t})\rangle=\mathcal{E}_{\lambda}(t;E_{t})|\lambda(t;E_{t})\rangle,
\label{eq:TD-Flo-func}
\end{equation}
with the Floquet Hamiltonian $\hat{H}_{F}(t;E_{t})\equiv\hat{H}_{I}(t;E_{t})-i\partial/\partial t$.

It is expedient to enumerate the Floquet quasienergy vectors as $|\kappa,l\rangle=|\kappa\rangle\bigotimes|l\rangle$, where $|\kappa\rangle$ is the system index, and $|l\rangle$ are the Fourier vectors. For each instant, we can write the time-independent generalized Floquet matrix eigenvalue equation as
\begin{equation}
\sum\limits_{\eta}\sum\limits_{l^{\prime}}\langle \kappa,l|\hat{H}_{F}(t;E_{t})|\eta,l^{\prime}\rangle\langle \eta,l^{\prime}|\lambda(t;E_{t})\rangle=\mathcal{E}_{\lambda}(t;E_{t})\langle\kappa,l|\lambda(t;E_{t})\rangle.
\label{eq:Flo-func}
\end{equation}
In this paper, we set $|l,l^{\prime}| \le 20$ to make sure the Floquet calculation is accessible and convergent. After solving the eigenvalue problem, we can obtain a set of quasienergy eigenvalues $\mathcal{E}_{\lambda}(t;E_{t})$ and corresponding eigenstates $|\lambda(t;E_{t})\rangle$ \cite{PRep390-1}. And the time-dependent wavefunction can be written as
\begin{equation}
|\Psi(t)\rangle=\sum\limits_{\lambda}a_{\lambda}(t)e^{-i\theta_{\lambda}(t;E_{t})}|\lambda(t;E_{t})\rangle,
\label{eq:TDWFd}
\end{equation}
where $a_{\lambda}$ are the probability amplitudes of the instantaneous Floquet quasienergy states. Substituting Eq.(\ref{eq:TDWFd}) into the TDSE with the Hamiltonian Eq.(\ref{eq:HI}), we obtain
\begin{equation}
\sum\limits_{\lambda}\left(\dot{a}_{\lambda}(t)|\lambda(t;E_{t})\rangle+a_{\lambda}(t)\dot{\varepsilon}_{t}\frac{\partial}{\partial E_{t}}|\lambda(t;E_{t})\rangle   \right)e^{-i\theta_{\lambda}(t;E_{t})}=0.
\label{eq:coeff1}
\end{equation}
Projecting onto an $|\lambda^{\prime}(t;E_{t})\rangle$, yields equations of motion for the coefficients,
\begin{equation}
\dot{a}_{\lambda^{\prime}}(t)=-a_{\lambda^{\prime}}(t)\dot{E_{t}}\langle \lambda^{\prime}|\frac{\partial}{\partial E_{t}}|\lambda^{\prime}\rangle+\sum\limits_{\lambda\neq \lambda^{\prime}}a_{\lambda}(t)\dot{E_{t}}\frac{  \langle \lambda^{\prime}|\partial \hat{H}_{F}/\partial E_{t} |\lambda\rangle e^{-i(\theta_{\lambda}-\theta_{\lambda^{\prime}})}  }{\mathcal{E}_{\lambda^{\prime}}-\mathcal{E}_{\lambda}},
\label{eq:coeff}
\end{equation}
in which $\langle \lambda^{\prime}|\partial/\partial E_{t}|\lambda\rangle=\langle \lambda^{\prime}|\partial \hat{H}_{I}/\partial E_{t} |\lambda\rangle/ (\mathcal{E}_{\lambda^{\prime}}-\mathcal{E}_{\lambda^{\prime}})$ is obtained by differentiating Eq.(\ref{eq:TD-Flo-func}) with respect to $E_{t}$, and the $t$ and $E_{t}$ dependence is suppressed for brevity. The time-dependent dipole moment related to $|\lambda_{\alpha}\rangle\approx |\alpha\rangle$ can be written as:
\begin{equation}
d(t)=a^{\ast}_{\alpha}(t)\sum_{l=-20}^{20}\sum_{\kappa=\beta,\gamma}a_{\kappa,l}(t)\langle \lambda_{\alpha}|\hat{z}|\lambda_{\kappa,l} \rangle e^{-i\theta_{\alpha\kappa}-il\omega_{I} t}+c.c.,
\label{eq:td_dipole_flo}
\end{equation}
with $\theta_{\alpha\kappa}=\int_{\tau}^{t}dt^{\prime}[\mathcal{E}_{\kappa}(t^{\prime})-\mathcal{E}_{\alpha}(t^{\prime})]$.

To explore the laser-induced lineshape change in the transient absorption spectra, we take the spectral line near the $1s^2-1s2p$ resonant absorption peak as an example, which corresponds to the time-dependent dipole moment between instantaneous Floquet quasienergy states $|\lambda_{\alpha}\rangle$ and $|\lambda_{\beta,0}\rangle$,
\begin{equation}
d_{\beta,0}(t)=a_{\beta,0}(t)\mu_{\beta\alpha}(t;E_{t}) e^{-i\theta_{\alpha\beta}}+c.c.,
\label{eq:td_dipole_flo2}
\end{equation}
in which $\mu_{\beta\alpha}(t;E_{t})=\langle \lambda_{\alpha} |\hat{z}|\lambda_{\beta,0} \rangle$ is the time-dependent transition matrix element, and $a_{\alpha}(t)\approx1$ is taken due to the weak XUV pulse. As we restrict our discussion to the situation that the XUV pulse precedes and does not overlap with the IR pulse, the coefficient  $a_{\beta,0}(t)$ can be obtained by numerically solving Eq.(\ref{eq:coeff}) with the initial condition
\begin{equation}
a_{\kappa,l}(t=0)=\left\{
{\begin{array}{ll}
ia_{X}&, \quad \text{for}\quad\kappa=\beta\ \text{and}\ l=0\\
0&, \quad \text{others},\\
\end{array}
}\right.
\label{eq: init}
\end{equation}
and $a_{X}$ being a real, time-independent constant. To first order in the XUV field $a_{X}=\mu^{0}_{\beta\alpha}E_{X}\tau_{X}$, where $E_{X}$ is the XUV field strength, and $\tau_{X}$ is a characteristic time for the XUV pulse, assumed to be much shorter than the period of IR field. We can recast the amplitude of the IR-dressed $1s2p$ state as
\begin{equation}
a_{\beta,0}(t)=ia_{X}a_{\delta}(t)e^{-i\theta_{\delta}(t)}
\label{eq:at_recast}
\end{equation}
with real time-dependent variables $a_{\delta}(t)$ and $\theta_{\delta}(t)$, decomposing the effect of the IR driving field on the $|\lambda_{\beta,0}\rangle$ state into the time-dependent population modulation $a_{\delta}(t)$ and the phase deviation $\theta_{\delta}(t)$. Then Eq.(\ref{eq:td_dipole_flo2}) can be rewritten as
\begin{equation}
d^{F}_{\beta,0}(t)=ia_{X}\mu_{\beta\alpha}a_{\delta}(t) e^{-i(\mathcal{E}^{0}_{\beta}-\mathcal{E}^{0}_{\alpha})t} e^{-i\theta_{T}(t)}+c.c.,
\label{eq:td_dipole_flo4}
\end{equation}
in which
\begin{equation}
\theta_{T}(t)=\theta_{\delta}(t)+\theta_{\alpha\beta}-(\mathcal{E}^{0}_{\beta}-\mathcal{E}^{0}_{\alpha})t
\label{eq:total_phase}
\end{equation}
is the total phase shift imported on the time-dependent dipole moment. Then the response function can be expressed, within the TDGF approach, as
\begin{equation}
S_{F}(\omega)=2\tilde{\varepsilon}_{X}\text{Im}[\tilde{d}_{F}(\omega)]
\label{eq:ResFun_GF}
\end{equation}
with
\begin{equation}
\label{eq:dip_F_omega}
\tilde{d}_{F}(\omega)=ia_{X}\int_{\tau}^{\infty}a_{\delta}(t) \mu_{\beta\alpha} e^{-i\theta_{T}(t)} e^{i[\omega-(\mathcal{E}^{0}_{\beta}-\mathcal{E}^{0}_{\alpha})]t} W(t)dt,
\end{equation}
by using the approximation that the broadband IAP spectral amplitude $\tilde{\varepsilon}_{X}$ is real and constant around $(\mathcal{E}^{0}_{\beta}-\mathcal{E}^{0}_{\alpha})$.

\subsection{The \emph{adiabatic} laser-induced phase and the \emph{non-adiabatic} phase correction}
\label{sec_LIP}
In the LIP model \cite{PRA88-033409}, Chen \emph{et al} assume that: the IR laser field does not modify the amplitude of the $1s2p$ state, but \emph{only} varies the $1s2p$ state energy and thus imposes the LIP, $\theta_{L}(t)=\int_{\tau}^{t}dt^{\prime}[\delta \mathcal{E}_{\beta}(t^{\prime})-\delta \mathcal{E}_{\alpha}(t^{\prime})]$ on the dipole moment with $\delta \mathcal{E}_{\alpha}(t)=\mathcal{E}_{\alpha}(t;E_{t})-\mathcal{E}^{0}_{\alpha}$ and $\delta \mathcal{E}_{\beta}(t)=\mathcal{E}_{\beta}(t;E_{t})-\mathcal{E}^{0}_{\beta}$ being the ac Stark shifts of $1s^2$ and $1s2p$ states, respectively. And the response function can be expressed as
\begin{equation}
S_{A}(\omega)=2\tilde{\varepsilon}_{X}\text{Im}[\tilde{d}_{A}(\omega)],
\label{eq:ResFun_A}
\end{equation}
with
\begin{equation}
\label{eq:dip_A_omega}
\tilde{d}_{A}(\omega)=ia_{X}\mu^{0}_{\beta\alpha}\int_{\tau}^{\infty} e^{-i\theta_{L}(t)} e^{i[\omega-(\mathcal{E}^{0}_{\beta}-\mathcal{E}^{0}_{\alpha})]t} W(t)dt.
\end{equation}

Note that, the LIP can also be derived from Eq.(\ref{eq:coeff}) by employing the generalized adiabatic theorem \cite{PRXQ2-030302}. If $E_{t}$ changes much more slowly than the difference in Floquet quasienergies, the slow modulation will not cross-couple the Floquet quasienergy states, i.e. the second term on the right of Eq.(\ref{eq:coeff}) can be ignored. Then $a_{\beta,0}(t)\approx ia_{X}$, $\theta_{\delta}(t)\approx 0$, and the time-dependent dipole moment between $|\lambda_{\alpha}\rangle$ and $|\lambda_{\beta,0}\rangle$ can be rewritten as
\begin{equation}
d^{A}_{\beta,0}(t)\approx ia_{X}\mu^{0}_{\beta\alpha} e^{-i(\mathcal{E}^{0}_{\beta}-\mathcal{E}^{0}_{\alpha})t} e^{-i\theta_{d}(t)}+c.c.
\label{eq:td_dipole_flo3}
\end{equation}
with the difference of dynamical phase shifts of Floquet quasienergy states
\begin{equation}
\begin{split}
\theta_{d}(t)=&\theta_{\alpha\beta}-(\mathcal{E}^{0}_{\beta}-\mathcal{E}^{0}_{\alpha})t\\
=&(\theta_{\beta}-\mathcal{E}^{0}_{\beta}t)-(\theta_{\alpha}-\mathcal{E}^{0}_{\alpha}t)\\
=&\int_{\tau}^{t}dt^{\prime}[\delta \mathcal{E}_{\beta}(t^{\prime})-\delta \mathcal{E}_{\alpha}(t^{\prime})],
\end{split}
\label{eq:dynamical_phase_shift}
\end{equation}
being the same as the LIP, $\theta_{L}(t)$. So one can consider the LIP model as the \emph{adiabatic} derivation by employing the GAA.

Consider the IR-driven dynamics in the Floquet picture. The Floquet quasienergy state $|\lambda_{\beta,0}\rangle$ under the GAA, namely the adiabatic Floquet state, evolves independently of other quasienergy states. The IR-driven coupling introduces a dynamical phase shift to the adiabatic Floquet state, importing the \emph{adiabatic} LIP, $\theta_{L}(=\theta_{d})$, on the time-dependent dipole moment. While the IR-driven cross-coupling between adiabatic Floquet states, i.e. the second term on the right of Eq.(\ref{eq:coeff}) that neglected in the GAA, introduces a \emph{non-adiabatic} phase correction, $\theta_{\delta}(t)$, to the adiabatic Floquet state.

 \section{results and discussions}
\label{sec_dis}
In this section, we numerically calculate the ATA spectra of a He-like three-level system by employing the TDGF approach, and investigate the laser-induced lineshape change in the frame of adiabatic Floquet states, via the comparison with the results based on the \emph{adiabatic} LIP model and RWA. Note that, in the TDGF calculation, i.e. Eqs.(\ref{eq:ResFun_GF}) and (\ref{eq:dip_F_omega}), only the time-dependent dipole moment between $|\lambda_{\beta,0}\rangle$ and $|\lambda_{\alpha}\rangle$ is considered. The results based on the \emph{adiabatic} LIP model are obtained by employing the GAA, Eqs.(\ref{eq:ResFun_A}) and (\ref{eq:dip_A_omega}). And the results based on the RWA are obtained from the first-order Floquet calculation, since the rotating-wave terms are dropped in the first-order Floquet theory \cite{JPB49-062003}. The parameters of the He-like three-level system are obtained via the diagonalization of the field-free Hamiltonian within the single-active-electron (SAE) approximation by employing the generalized pseudospectral method (see the Appendix for details), which gives $\mathcal{E}^{0}_{\alpha}=-0.90369$a.u.($-24.59$eV), $\mathcal{E}^{0}_{\beta}=-0.12384$a.u.($-3.37$eV), $\mathcal{E}^{0}_{\gamma}=-0.14589$a.u.($-3.97$eV), $\mu_{\alpha\beta}=0.2084$a.u., and $\mu_{\beta\gamma}=2.9471$a.u.. A $330$as-FWHM Gaussian XUV pulse with carrier frequency $\omega_{X}=25$eV and peak intensity $1\times10^{10}$W cm$^{-2}$, and a $40$fs-delayed $15$fs-FWHM Gaussian IR pulse are employed. The carrier frequency and peak intensity of the IR pulse are set for the following specific cases.

\subsection{Weak driving and moderate detuning}
\label{subsec_wm}

\begin{figure}[ht]
\includegraphics[width=\columnwidth]{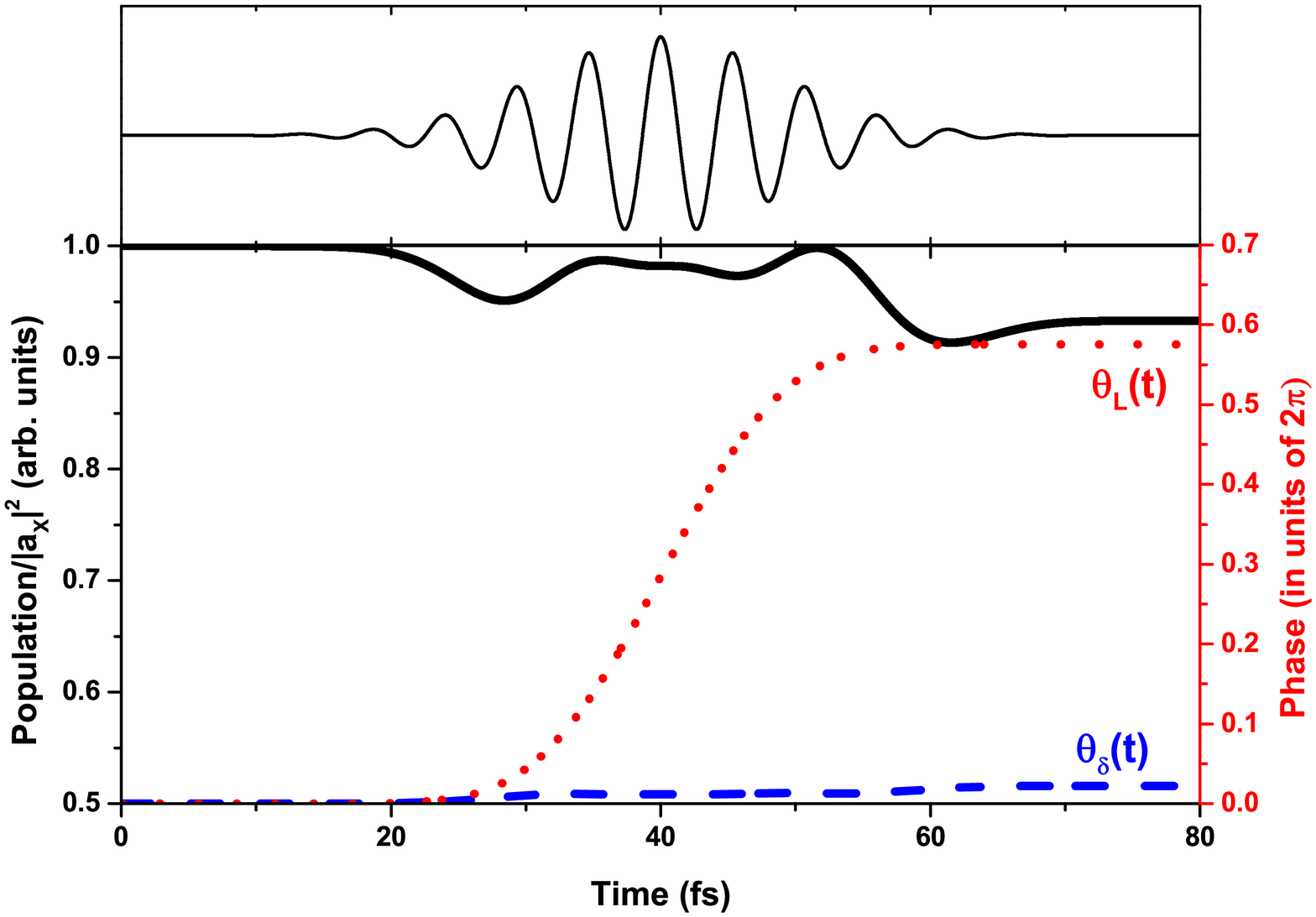}
\caption{The evolution of the Floquet quasienergy state $|\lambda_{\beta,0}(t)\rangle$ based on the TDGF approach, in the case of weak driving and moderate detuning, $I_{IR}=1\times 10^{12}$W/cm$^{2}$ and $\omega_{I}=0.77$eV. The normalized population $|a_{\beta,0}|^2/|a_{X}|^2$ is represented with black-solid line. The adiabatic LIP and the non-adiabatic phase correction are represented with red-dotted and blue-dash lines, respectively. The top panel illustrates the IR pulse centered at $\tau=40$fs.}
\label{fig:0.77c}
\end{figure}

\begin{figure}[ht]
\includegraphics[width=\columnwidth]{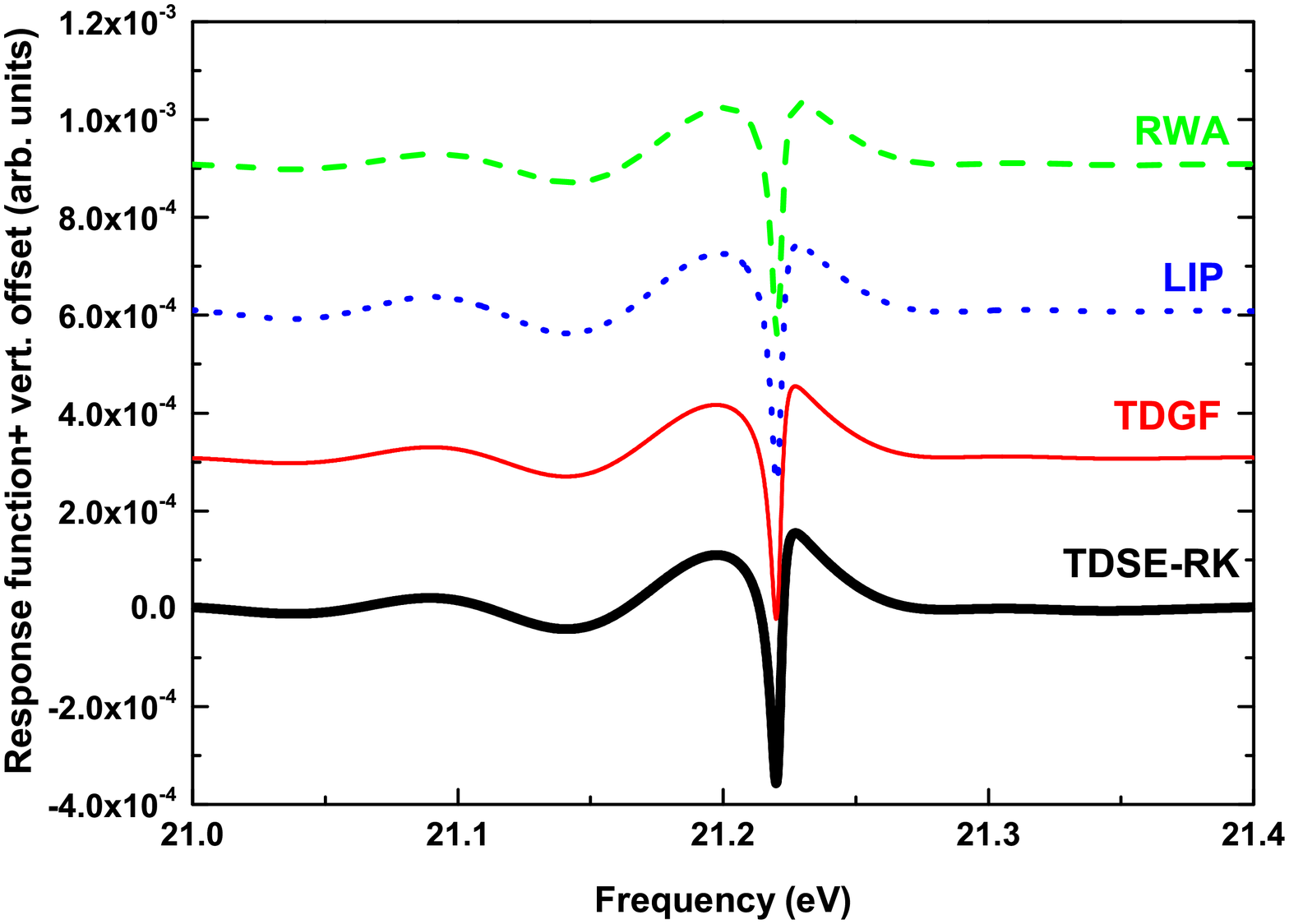}
\caption{The response functions calculated from different approaches are shown in the region of the $1s^{2}$-$1s2p$ resonant absorption peak, in the case of weak driving and moderate detuning. The black-thick-solid, red-thin-solid, blue-dotted and green-dash lines are based on the numerical RK-method, the TDGF approach, the LIP model and the RWA, respectively.}
\label{fig:0.77s}
\end{figure}

In the case of weak driving and moderate detuning, we set the peak intensity of the IR pulse $I_{IR}=1\times 10^{12}$W/cm$^{2}$ and carrier frequency $\omega_{I}=0.77$eV. The probability amplitude of the Floquet quasienergy state $|\lambda_{\beta,0}\rangle$ is calculated by employing the TDGF approach. In Fig.\ref{fig:0.77c}, we plot the normalized population $|a_{\beta,0}(t)|^2/|a_{X}|^2$ (black-solid line), the LIP $\theta_{L}(t)$ (red-dotted line) and the non-adiabatic phase correction $\theta_{\delta}(t)$ (red-dash line) as a function of time. The population of $|\lambda_{\beta,0}\rangle$ is modulated due to the IR-induced cross-coupling with other adiabatic Floquet states, while $\theta_{L}(t)$ and $\theta_{\delta}(t)$ are accumulated during the IR-atom interaction. $|a_{\beta,0}(t)|^2/|a_{X}|^2\approx1$ implies that the IR-induced coupling between $|\lambda_{\beta,0}\rangle$ and other adiabatic Floquet states can be ignored, and $\theta_{\delta}(t)\ll \theta_{L}(t)$ leads to $\theta_{\delta}(t)\approx 0$. In other words, the GAA is a good approximation. It is reasonable to consider the evolution of $|\lambda_{\beta,0}\rangle$ adiabatically, so one might expect the TDGF calculation would agree well with the TDSE-RK calculation with Eq.(\ref{eq:ResFun}). Meanwhile the GAA indicates that the resonant absorption lineshape based on the \emph{adiabatic} LIP model would agree well with that based on the TDGF approach. Indeed, in Fig.\ref{fig:0.77s} we see good agreement of the TDGF calculation with the TDSE-RK calculation, as well as with the \emph{adiabatic} LIP model. Fig.\ref{fig:0.77s} also present the result obtained by the RWA that fits with the above three calculation well, implying that the contribution of the rotating-wave terms, i.e. the higher-order Floquet quasienergy vectors, are neglectable in the adiabtic Floquet state $|\lambda_{\beta,0}\rangle$.

\subsection{Strong driving and large detuning}
\label{subsec_sl}

\begin{figure}[ht]
\includegraphics[width=\columnwidth]{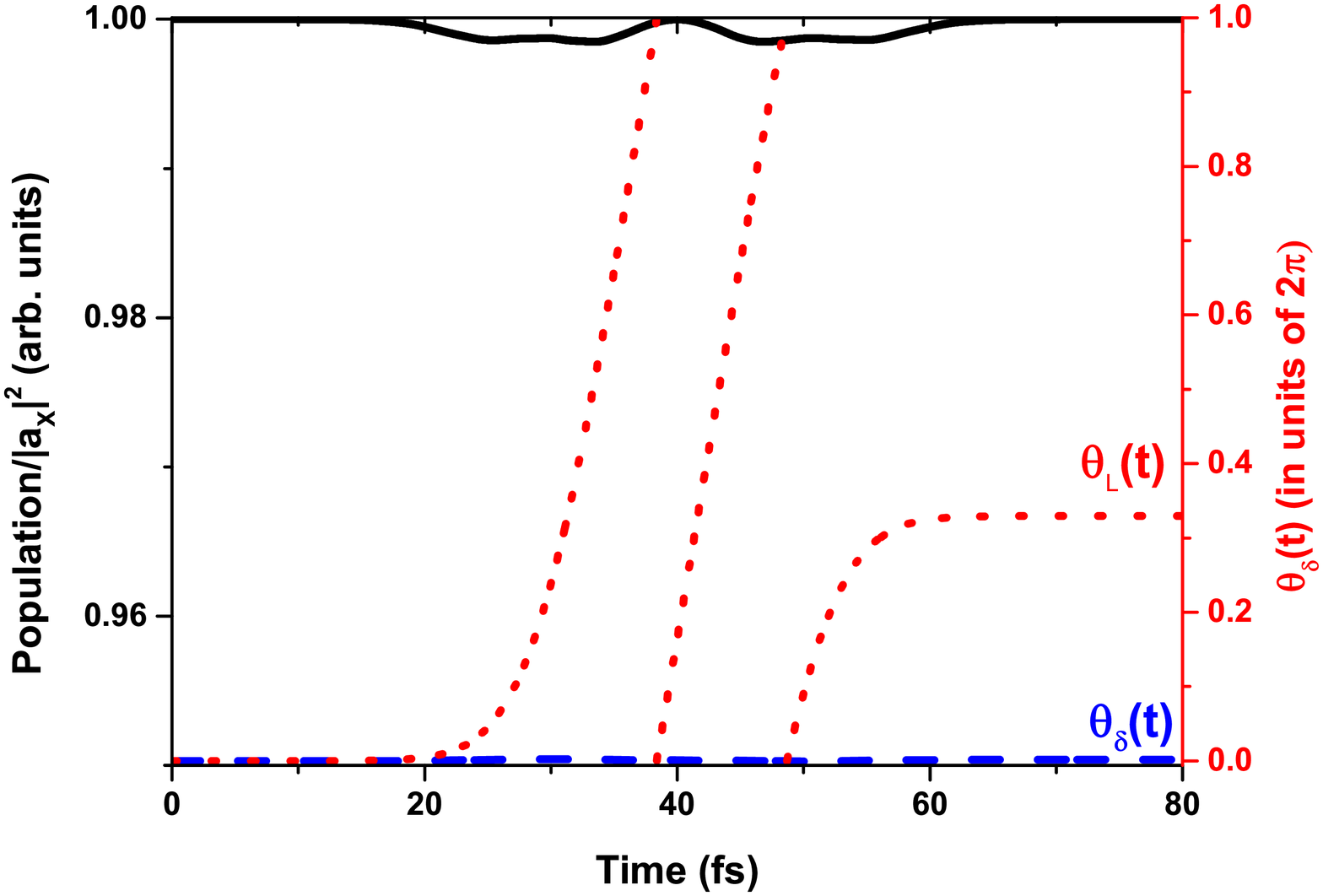}
\caption{The evolution of the Floquet quasienergy state $|\lambda_{\beta,0}(t)\rangle$ based on the TDGF approach, in the case of strong driving and large detuning, $I_{IR}=5\times 10^{13}$W/cm$^{2}$ and $\omega_{I}=1.55$eV. The normalized population $|a_{\beta,0}|^2/|a_{X}|^2$ is represented with black solid line. The adiabatic LIP and the non-adiabatic phase correction are represented with red-dotted and blue-dash lines, respectively.}
\label{fig:1.55c}
\end{figure}

\begin{figure}[ht]
\includegraphics[width=\columnwidth]{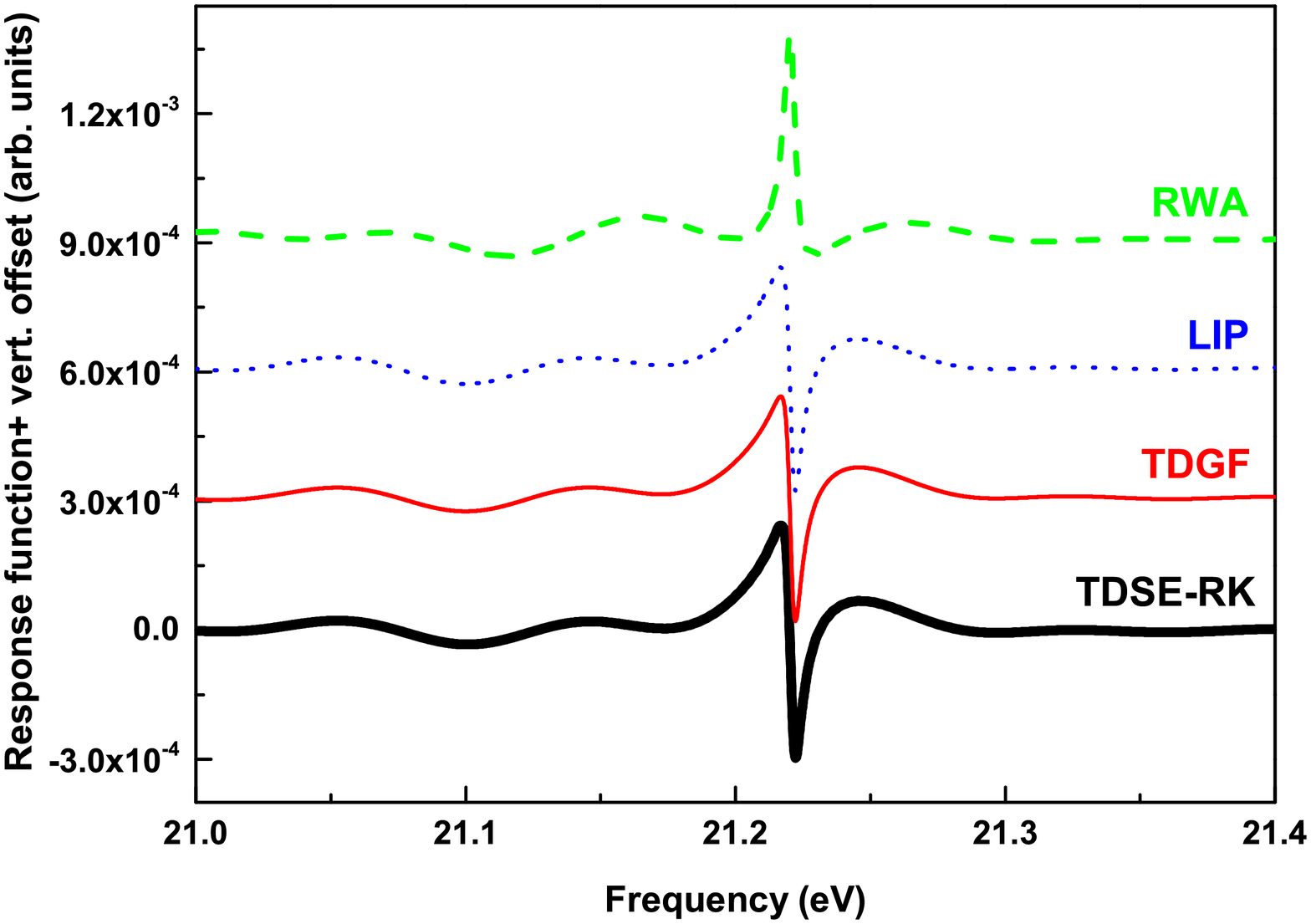}
\caption{The response functions calculated from different approaches are shown in the region of the $1s^{2}$-$1s2p$ resonant absorption peak, in the case of strong driving and large detuning. The black-thick-solid, red-thin-solid, blue-dotted and green-dash lines are based on the numerical RK-method, the TDGF approach, the LIP model and the RWA, respectively.}
\label{fig:1.55s}
\end{figure}

\begin{figure}[ht]
\includegraphics[width=\columnwidth]{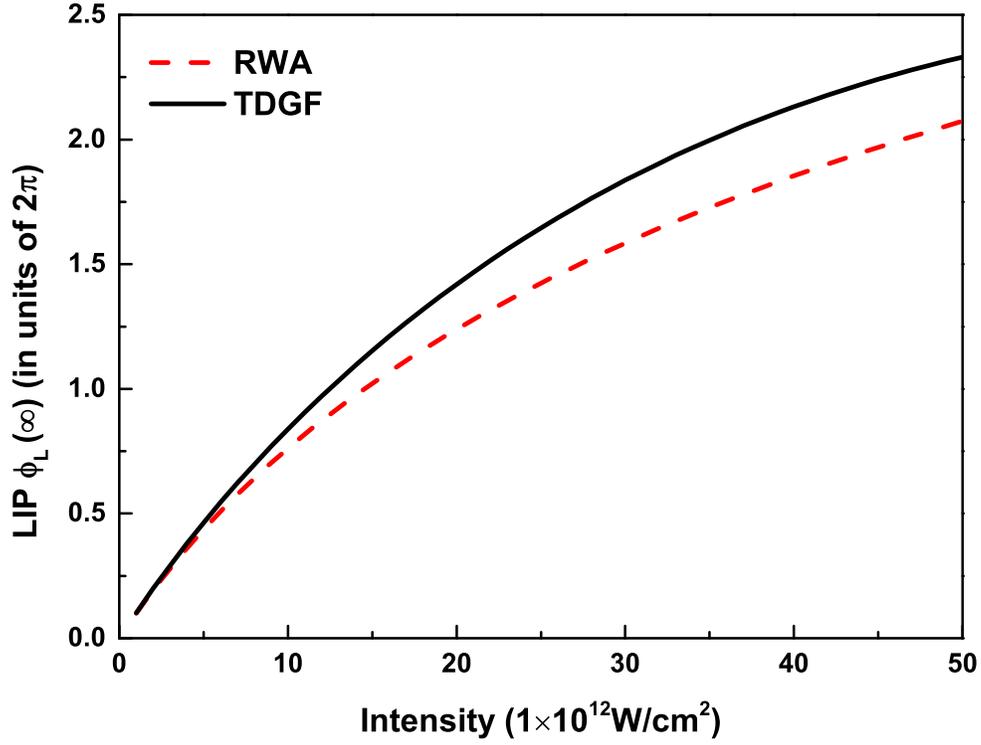}
\caption{The dependence of the LIP on the driving field intensity calculated from the RWA (black-solid line) and the TDGF approach (red-dotted line), respectively, in the case of large detuning $\omega_{I}=1.55$eV.}
\label{fig:1.55Pt}
\end{figure}

In the case of strong driving and large detuning, we set $I_{IR}=5\times 10^{13}$W/cm$^{2}$ and $\omega_{I}=1.55$eV. In Fig.\ref{fig:1.55c}, we plot the normalized population $|a_{\beta,0}(t)|^2/|a_{X}|^2$ (black-solid line), the LIP $\theta_{L}(t)$ (red-dash line) and the non-adiabatic phase correction $\theta_{\delta}(t)$ (blue-dotted line), as a function of time. Similar to the case of weak driving and moderate detuning, $|\lambda_{\beta,0}\rangle$ evolves adiabatically and the GAA is a good approximation. Fig.\ref{fig:1.55s} shows that, as expected, the TDGF approach and the adiabatic LIP model agree well with the TDSE-RK calculation. However, the RWA fails to reproduce the laser-induced change of the resonant absorption lineshape.  As $\theta_{\delta}(t)\ll \theta_{L}(t)$, the reasonable deduction is that the \emph{adiabatic} LIP imported on the time-dependent dipole moment can not be calculated accurately by the RWA. We plot the LIPs after the IR driving pulse as a function of the peak intensity in Fig.\ref{fig:1.55Pt}. It is shown that, the weaker the peak intensity of the IR driving pulse is, the better the LIP calculated by the RWA agrees with that calculated by the TDGF approach. It is straightforward to deduce that the breakdown of RWA is due to the dropped rotating-wave terms. This can be understood in the frame of adiabatic Floquet states. As the driving gets stronger, higher-order Floquet quasienergy vectors are coupled in the formation of the adiabatic Floquet state $|\lambda_{\beta,0}\rangle$. As a result, the RWA, i.e. the 'first-order' Floquet calculation, cannot give the convergent adiabatic Floquet states and their eigenenergies, leading to incorrect LIP obtained at higher peak intensity of the IR pulse.

\subsection{Weak driving and small detuning}
\label{subsec_ws}

\begin{figure}[ht]
\centering
\begin{minipage}[c]{0.5\textwidth}
\centering
\includegraphics[width=\columnwidth]{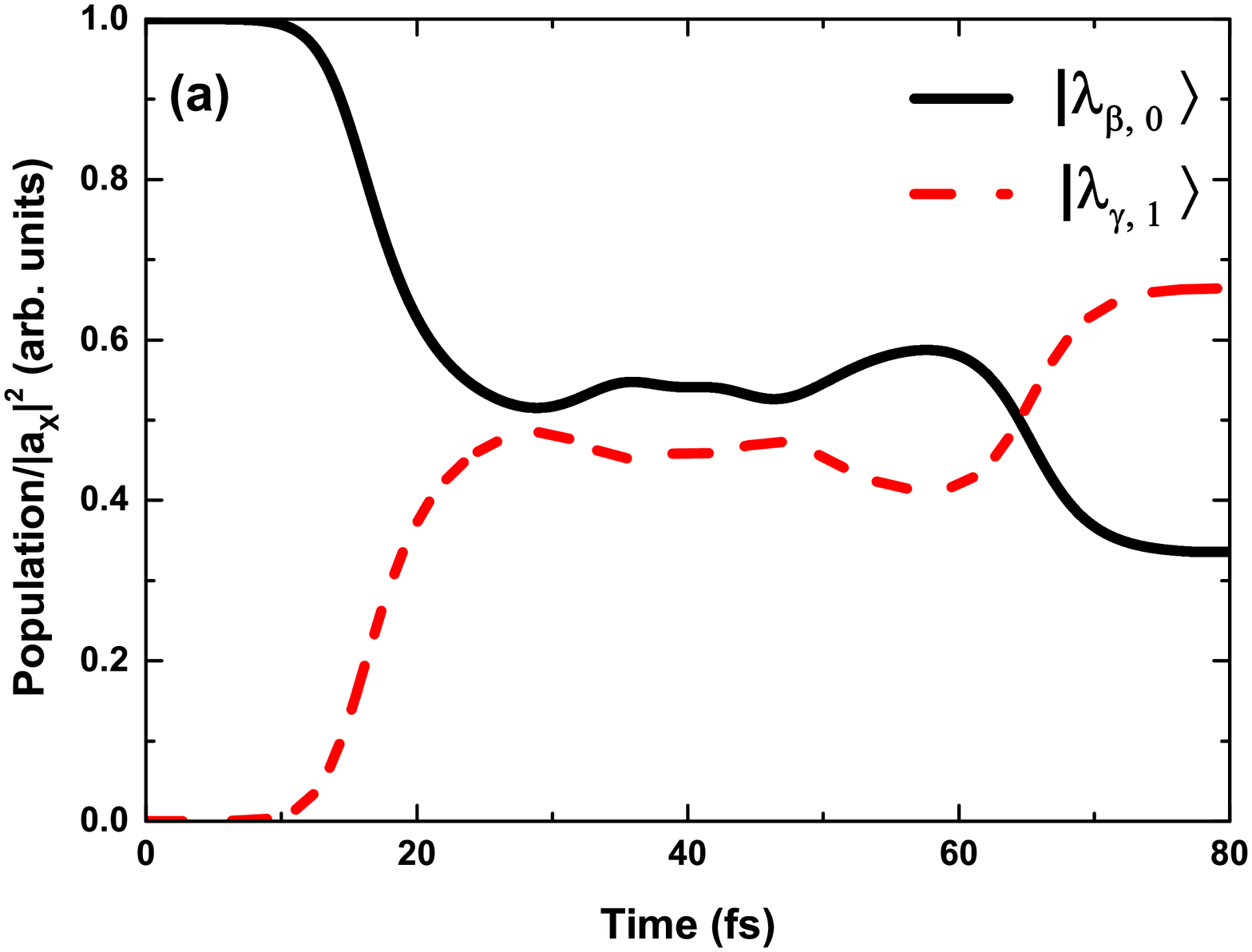}
\end{minipage}%
\begin{minipage}[c]{0.5\textwidth}
\centering
\includegraphics[width=\columnwidth]{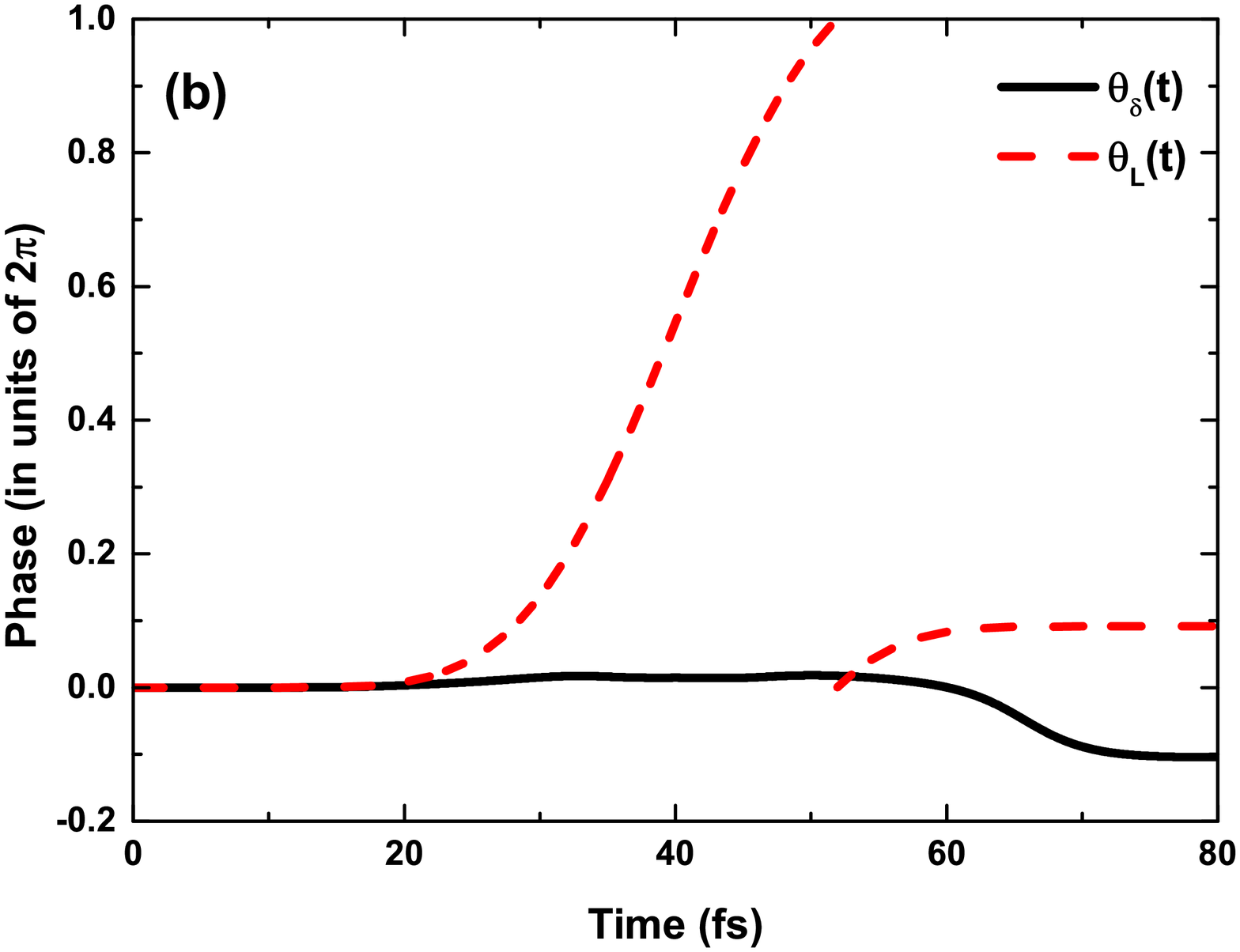}
\end{minipage}
\caption{In the case of weak driving and small detuning, (a) the normalized population of the Floquet quasienergy states, $|\lambda_{\beta,0}\rangle$ and $|\lambda_{\gamma,1}\rangle$, are plotted with black-solid and red-dash lines, respectively. (b) The LIP and the non-adiabatic phase correction of the Floquet quasienergy state $|\lambda_{\beta,0}(t)\rangle$ are plotted with red-dash and black-solid lines, respectively.}
\label{fig:0.61cp}
\end{figure}

\begin{figure}[ht]
\includegraphics[width=\columnwidth]{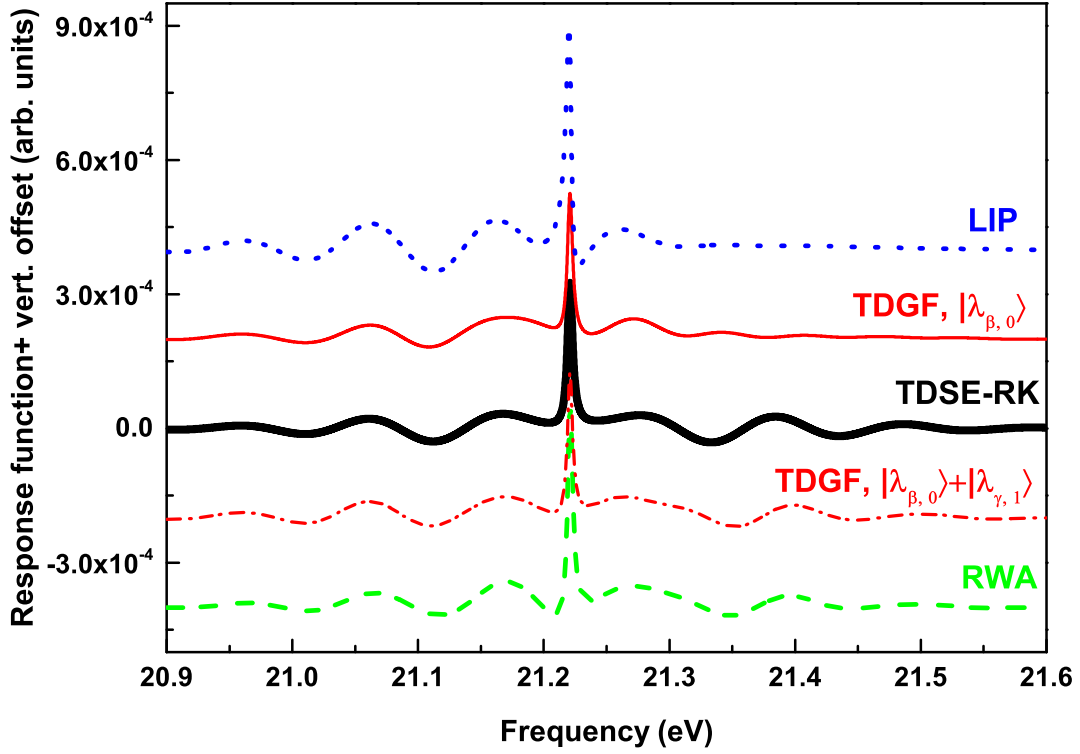}
\caption{In the case of weak driving and small detuning, $I_{IR}=1\times 10^{12}$W/cm$^{2}$ and $\omega_{I}=0.61$eV, the response functions calculated from different approaches are shown in the region of the $1s^{2}$-$1s2p$ resonant absorption peak. The black-thick-solid, red-thin-solid and blue-dotted lines are based on the numerical RK-method, the TDGF approach and the LIP model, respectively. The green-dash and red-dash-dotted lines are based on the RWA and the improved TDGF approach, respectively, including Floquet quasienergy states $|\lambda_{\beta,0}\rangle$ and $|\lambda_{\gamma,1}\rangle$.}
\label{fig:0.61s1}
\end{figure}

In the case of weak driving and small detuning, we set $I_{IR}=1\times 10^{12}$W/cm$^{2}$ and $\omega_{I}=0.61$eV.
We plot the normalized population $|a_{\beta,0}(t)|^2/|a_{X}|^2$ with black-solid line in Fig.\ref{fig:0.61cp}(a), and plot the LIP $\theta_{L}(t)$ (black-solid line) and the non-adiabatic phase correction $\theta_{\delta}(t)$ (red-dash line) in Fig.\ref{fig:0.61cp}(b), respectively. It is shown that, the evolution of $|\lambda_{\beta,0}\rangle$ is non-adiabatic, and the non-adiabatic phase correction is non-negligible. Thus the GAA breaks down, and the calculation based on the \emph{adiabatic} LIP model can not reproduce the spectral lineshape well, as presented in Fig.\ref{fig:0.61s1}. It is also shown that, the calculation based on the TDGF approach reproduces the spectral lineshape in the TDSE-RK calculation well in the region of the $1s^2-1s2p$ resonant absorption peak and the low-frequency sideband, but not in the region of the high-frequency sideband. As only the dipole transition between $|\lambda_{\beta,0}\rangle$ and $|\lambda_{\alpha,0}\rangle$ is included in the Floquet calculation, i.e. Eqs.(\ref{eq:td_dipole_flo2}) and (\ref{eq:dip_F_omega}), one can deduce that this disagreement in the high-frequency sideband is due to the neglection of Floquet states that populated by the IR pulse. Our TDGF calculation indicates that, non-negligible population is excited from $|\lambda_{\beta,0}\rangle$ to $|\lambda_{\gamma,1}\rangle$ owing to the non-adiabatic near-resonant coupling driven by the IR pulse, as shown with red-dash line in Fig.\ref{fig:0.61cp}(a). We rewrite the Eqs.(\ref{eq:ResFun_GF}) and (\ref{eq:dip_F_omega}), to include the dipole transition between $|\lambda_{\gamma,1}\rangle$ and $|\lambda_{\alpha,0}\rangle$, as
\begin{equation}
S^{\prime}_{F}(\omega)=2\tilde{\varepsilon}_{X}\text{Im}[\tilde{d}^{\prime}_{F}(\omega)]
\label{eq:ResFun_GF2}
\end{equation}
with
\begin{equation}
\label{eq:dip_F_omega2}
\tilde{d}^{\prime}_{F}(\omega)=ia_{X} \int_{\tau}^{\infty}\Big[a_{\delta}(t) \mu_{\beta\alpha}e^{-i\theta_{T}(t)} e^{i[\omega-(\mathcal{E}^{0}_{\beta}-\mathcal{E}^{0}_{\alpha})]t}+a_{\gamma,1}(t) \mu_{\gamma\alpha} \Big]W(t)dt
\end{equation}
in which $a_{\gamma,1}(t)$ is the probability amplitude of $|\lambda_{\gamma,1}\rangle$, and $\mu_{\gamma\alpha}(t;E_{t})=\langle \lambda_{\alpha} |\hat{z}|\lambda_{\gamma,1} \rangle$ is the transition matrix element between $|\lambda_{\gamma,1}\rangle$ and $|\lambda_{\alpha,0}\rangle$. With the improved version of TDGF calculation, Eqs.(\ref{eq:ResFun_GF2}) and (\ref{eq:dip_F_omega2}), we obtain the spectral lineshape which agrees well with the TDSE-RK calculation in the regions of both the absorption peak and sidebands, as presented with red-dash-dotted line in Fig.\ref{fig:0.61s1}. It is also presented that, the result based on the RWA calculation including both $|\lambda_{\beta,0}\rangle$ and $|\lambda_{\gamma,1}\rangle$ agrees well with both the TDSE-RK and the improved TDGF calculations. It is implied that the first-order Floquet quasienergy vectors are the major components of the Floquet states $|\lambda_{\beta,0}\rangle$ and $|\lambda_{\gamma,1}\rangle$, and the \emph{non-adiabatic} near-resonant IR-coupling between adiabatic Floquet states is involved in the RWA.

\subsection{Comparison with \emph{ab initio} calculation}
\label{subsec_ab}

\begin{figure}[ht]
\includegraphics[width=0.75\columnwidth]{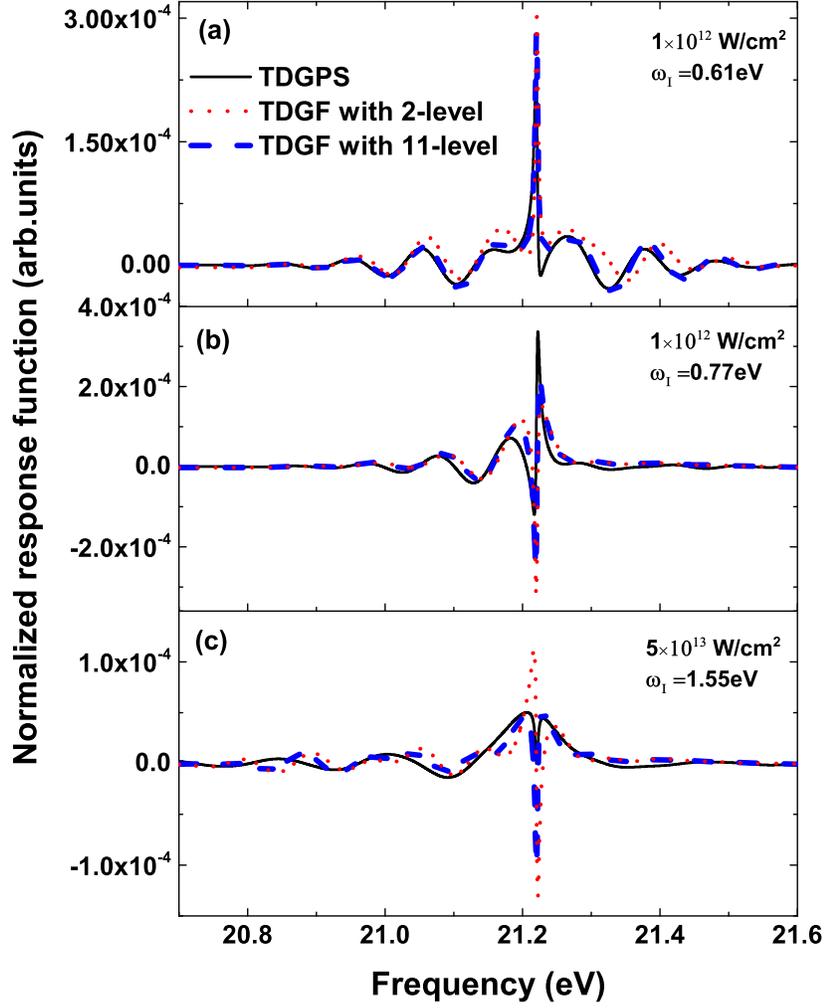}
\caption{The comparison of response functions based on different approaches in the cases with (a) $I_{IR}=1\times 10^{12}$W/cm$^{2}$ and $\omega_{I}=0.77$eV, (b) $I_{IR}=1\times 10^{12}$W/cm$^{2}$ and $\omega_{I}=0.61$eV and (c) $I_{IR}=5\times 10^{13}$W/cm$^{2}$ and $\omega_{I}=1.55$eV. The TDGPS calculation is plotted with black-solid line, and the TDGF calculations with 2-level and 11-level models are plotted with red-dotted and blue-dash lines, respectively. }
\label{fig:GPSc}
\end{figure}

Although the above investigations are just with primitive two-level model, we confirm the efficiency of the TDGF approach via the comparison with the \emph{ab initio} calculation calculation of SAE Helium atom (see the Appendix for details) in the three cases. As presented by Fig.\ref{fig:GPSc}, the TDGF calculation (red-dotted line) is different from the \emph{ab initio} calculation (black solid). The difference is due to the ionization and the multi-level effect in real atomic systems. The multi-level effect can be improved by including more levels in the TDGF calculation. The TDGF calculation with 11-level model, including $1s2s-1s4s$, $1s2p-1s5p$ and $1s3d-1s6d$, is plotted in Fig.\ref{fig:GPSc} with blue-dash line,  presenting quite good agreement with the \emph{ab initio} calculation.

\section{Conclusion}
\label{sec_con}

In this paper, we introduce an alternative numerical approach, the time-dependent generalized Floquet (TDGF) approach, to calculate the ATA spectra of a He-like three-level system driven by the combination of an attosecond XUV pulse and a delayed few-cycle IR pulse. The main features of ATA spectra can be reproduced, though only the primitive two-level system involving the two upper levels and the IR pulse arriving after the XUV pulse is considered, and much better agreement with the \emph{ab initio} calculation can be achieved by including more levels.

In terms of adiabatic Floquet states, we derive the total phase shift imported on the time-dependent dipole moment, and demonstrate the adiabatic LIP and the non-adiabatic phase correction, by employing the generalized adiabatic theorem. The comparisons with the RWA and the \emph{adiabatic} LIP model are presented in three typical cases. All three approaches agree well with the numerical RK calculation in the case of weak driving and moderate detuning. The \emph{adiabatic} LIP model fails the case of weak driving and small detuning, due to the non-adiabatic near-resonant cross-coupling between adiabatic Floquet states. While the RWA falls the case of strong driving and large detuning, because the higher-order Floquet quasienergy vectors that are coupled in the formation of adiabatic Floquet states, i.e. the rotating-wave terms, are dropped off in the RWA.

In summary, the TDGF approach is an alternative numerical approach to investigate the transient dynamics driven by laser fields, providing an additional understanding of the laser induced lineshape change in ATA spectra in the frame of adiabatic Floquet states.

\acknowledgments
This work was supported by the Natural Science Foundation of China (Grants 12074307, 91536115, 11534008, 11504288) and the Natural Science Foundation of Shannxi Province (Grant 2019JM-410).

\appendix
\section{The \emph{ab initio} simulation}
Under the SAE approximation, the TDSE describing the interaction of a single He atom with laser field can be written as
\begin{equation}
i\frac{\partial}{\partial t}|\Psi(\bm{r},t)\rangle=\hat{H}(t)|\Psi(\bm{r},t)\rangle=[\hat{H}_{0}+\hat{V}(t)]|\Psi(\bm{r},t)\rangle.
\label{eq:TDSE}
\end{equation}
Here the laser-atom interaction within the dipole approximation is given by
\begin{equation}
\hat{V}(t)=-\bm{E}(t)\cdot\bm{r},
\end{equation}
and $\hat{H}_{0}$ represents the unperturbed atom Hamiltonian, which can be written as
\begin{equation}
\hat{H}_{0}=-\frac{1}{2}\bigtriangledown^{2}+\sum\limits_{l}|Y^{0}_{l}\rangle V_{l}\langle Y^{0}_{l}|,
\label{eq:GPS_H_0}
\end{equation}
where $V_{l}$ is model potential for He atom, and $Y^{0}_{l}$ is a spherical harmonic. We employ an angular-momentum-dependent model potential in the following form:
\begin{equation}
\begin{split}
V_{l}=&-\frac{\alpha}{2r^{4}}W_{6}\Big(\frac{r}{r_{c}}\Big)-\frac{1}{r}-\Big(\frac{N-S}{r}+A_{1}\Big)e^{-B_{1}r}\\
&-\Big(\frac{S}{r}+A_{2}\Big)e^{-B_{2}r},
\end{split}
\label{eq:model_pot}
\end{equation}
where $\alpha$ is the He$^{+}$ core dipole polarizability \cite{PRA89-023431,PRA64-013406}. $W_{6}$ is a core cutoff function given bey
\begin{equation}
W_{n}(x)=1-\Big[1+nx+\frac{(nx)^2}{2!}+\cdots+\frac{(nx)^2}{n!}\Big]e^{-nx},
\end{equation}
and $r_{c}$ is an effective He$^{+}$ core radius. The values of the parameters determined are listed in Table \ref{tab:MPP}. In the present work, it is sufficient to use four different angular-momentum-dependent model potentials.

\begin{table}
\centering
\caption{Model potential parameters for He (in a.u.).}
\begin{tabular}{@{\extracolsep{4pt}}c c c c c c c c}
\hline
\hline
l&$\alpha$&$r_{c}$&$S$&$A_{1}$&$A_{2}$&$B_{1}$&$B_{2}$\\
\hline
0&0.28125&2.0&-7.9093912&-10.899664&0.0&1.7&3.8\\
1&0.28125&2.0&1.50094970&0.11297684&0.0&1.3&3.8\\
2&0.28125&2.0&0.88294766&-0.032043029&0.0&1.3&3.8\\
3&0.28125&2.0&0.41193110&-0.129391180&0.0&1.3&3.8\\
$>$3&0.28125&2.0&0.0&0.0&0.0&1.3&0.0\\
\hline
\hline
\end{tabular}
\label{tab:MPP}
\end{table}

Via the diagonalization of the field-free Hamiltonian $\hat{H}_{0}$, one can obtain the eigenstates of the SAE He atom, and the transition matrix elements between them. The TDSE can be solved numerically by means of the time-dependent generalized pseudospectral (TDGPS) method \cite{chemphys217-119}. The maximum radius of the grid is set at $r_{max} = 150$, and the number of collocation points along the radial coordinate is $N_{r} =300$. The absorbing boundary starting at $r_{b} = 100$ is employed to avoid the artificial reflection. The pulse-envelope function is Gaussian shaped and centered around $t = 0$ for the IR and $\tau$ for the XUV, respectively. After the time propagation of the wavefunction $\Psi(\bm{r},t)$, we can calculate the expectation value of the induced dipole moment in the length,
\begin{equation}
\label{eq:dip_GPS}
d(t)=\langle\Psi(\bm{r},t)|z|\Psi(\bm{r},t)\rangle.
\end{equation}
And the absorption spectra can be obtained via substituting the Fourier transformation of Eq.(\ref{eq:dip_GPS}) multiplied with the decaying function Eq.(\ref{eq:winfunc}) into the response function Eq.(\ref{eq:ResFun}).


\end{document}